\documentclass[letterpaper, 10 pt, conference]{ieeeconf}
\usepackage{cite,amsmath,amssymb,amsfonts,algorithmic,graphicx,textcomp,xcolor,tikz,bm,adjustbox}
\IEEEoverridecommandlockouts
\usetikzlibrary{
  arrows.meta,
  positioning,
  shapes.geometric,
  fit,
  backgrounds,
  calc
}
\def\BibTeX{{\rm B\kern-.05em{\sc i\kern-.025em b}\kern-.08em
    T\kern-.1667em\lower.7ex\hbox{E}\kern-.125emX}}

\definecolor{DarkRed}{RGB}{180,30,40}
\definecolor{DeepBlue}{RGB}{0,70,160}
\definecolor{DarkGreen}{RGB}{0,130,70}
\definecolor{Violet}{RGB}{120,0,140}
\definecolor{BurntOrange}{RGB}{200,100,0}
\definecolor{TealBlue}{RGB}{0,120,140}

\begin{document}
\bstctlcite{IEEEexample:BSTcontrol}
\pagestyle{plain}

\title{Experimental validation of a fast control-oriented, physics-informed surrogate model for plasma equilibrium reconstruction in the TCV tokamak}
 
\author{M. Grandin$^{1}$,
A. Mele$^{2}$,
C. Galperti$^{2}$,
D. Gonzales Castineiras$^{2}$,
C. Hei{\ss}$^{2}$,
A. Cenedese$^{1}$, \\
TCV team$^{3}$
and EUROfusion Tokamak Exploitation team$^{4}$%
\thanks{$^{1}$University of Padova, Padova, Italy}%
\thanks{$^{2}$Ecole Polytechnique Federale de Lausanne (EPFL), Swiss Plasma Center (SPC), 1015 Lausanne, Switzerland}%
\thanks{$^{3}$See author list of C. Theiler et al. 2026 \textit{Nucl. Fusion} 66 116007}%
\thanks{$^{4}$See author list of E. Joffrin et al 2024 \textit{Nucl. Fusion} 64 112019}%
}

\maketitle

\begin{abstract}
    Magnetic equilibrium reconstruction provides the plasma state estimate required for real-time shape control in tokamaks. We present a fast, physics-informed neural network surrogate of the \texttt{liuqe} equilibrium reconstruction code \cite{liuqe1} for the TCV tokamak at EPFL, achieving inference times below 100~$\bm\mu$s and enabling 10~kHz shape control.

    The model is trained on around 10,000 TCV discharges spanning the full operational range of plasma shapes. Its modular branch/trunk architecture decouples magnetic measurement encoding from spatial coordinate processing, enabling physics-informed regularization via automatic differentiation of the predicted flux map.

    The surrogate has been compiled and deployed on the TCV real-time control system, and validated both offline and in real time against the models \texttt{liuqe-rt} and \texttt{lih}, showing comparable accuracy. Closed-loop performance assessed with the real-time software in-the-loop \texttt{fge} \cite{fge1} demonstrates control-equivalent behavior across multiple control strategies.
\end{abstract}

\section{Introduction}
\label{sec:intro}

Plasma shape control is a central challenge in tokamak operation, as the plasma boundary determines interactions with plasma-facing components, MHD stability margins, and confinement performance.
High control bandwidth is particularly important for advanced or inherently unstable plasma configurations, during fast current ramp-up and ramp-down transients, and more generally to improve shape-tracking accuracy.
A 10~kHz shape-control loop would be desirable on these grounds alone; it would additionally enable the unification of shape control with vertical position stabilization, which on most tokamaks already runs at high bandwidth.

The Tokamak à Configuration Variable (TCV) at EPFL is designed to explore a wide range of plasma shapes and control strategies \cite{theiler2026progress,hofmann1994creation}, being equipped with poloidal field (PF) coils and magnetic diagnostics (flux loops and field probes) around the vacuum vessel.
The standard equilibrium reconstruction code is \textsc{liuqe} \cite{liuqe1}, which solves the Grad--Shafranov equation (GSE) iteratively on a $28\times65$ grid.
\textsc{liuqe-rt} limits iterations and interpolation complexity, achieving cycle times of $\lesssim 0.5$~ms and sustaining the current 1~kHz shape-control loop; the vertical stabilization loop, by contrast, already runs at~10~kHz.
\textsc{LIH}, the zeroth-iteration first guess of the algorithm, operates at~10~kHz but with substantially reduced accuracy.
The 1~kHz constraint imposed by \textsc{liuqe-rt} is therefore the primary bottleneck preventing both higher-bandwidth shape control and unified shape-and-position control on TCV.

The underlying physical problem is the GSE \cite{GS1,GS2}, describing axisymmetric MHD force balance.
It is useful to distinguish three related but distinct GSE-constrained problems that appear in the literature.
The \emph{free-boundary} problem takes the currents in the PF coils and passive conductors as inputs and finds the flux map together with the plasma boundary.
The \emph{fixed-boundary} problem takes a prescribed last-closed flux surface (LCFS) as a constraint and finds the PF coil currents required to sustain it.
\emph{Equilibrium reconstruction}, the problem addressed in this paper, is an inverse problem: given a set of experimental measurements (magnetic flux loops and field probes around the vacuum vessel), it seeks the flux map~$\psi(r,z)$ and the two free functions~$p'(\psi)$ and~$TT'(\psi)$ that simultaneously satisfy the GSE and best reproduce those measurements in a least-squares sense.
In general, solving the inverse problem is computationally expensive and time-consuming, thereby motivating the development of fast and parsimonious surrogate models.

Neural networks (NN) surrogates for plasma equilibrium have attracted broad interest, spanning equilibrium reconstruction \cite{nneq1,nneq3,nneq4,nneq5,nneq6,nneq7_hyb,5svd_nn22,nneq12lehigh,nneq10}, free-boundary equilibrium \cite{nneq2}, and boundary and LCFS reconstruction \cite{caronte25pi,lcfs25,nn_fast_topology}, with many adopting a GSE-related physics-informed loss \cite{nneq4,nneq8pi23,caronte25pi,nneq11mix,bonotto24}.
A distinct subset uses Neural Operators \cite{DeepONet}, which learn function-space mappings by separating input-function encoding from spatial coordinate processing \cite{uniapprox95}, naturally yielding Physics-Informed Neural Operators (PINOs) via automatic differentiation.
Recent works have applied PINO-type architectures directly to the GSE \cite{gse_pinn1,gse_pinn2,gse_pinn3}, including in the fixed-boundary setting \cite{gse_pinn1}.

On the control side, the use of a fast NN as a shape-control observer was pioneered by Bishop et al.\ \cite{based95} ($<50$~$\mu$s with mixed digital--analog hardware) and Jeon et al.\ \cite{Jeon2001_nn_pos_ctrl}, and more recently demonstrated on Globus-M/M2 \cite{nt21eqr_ctrl}, albeit with a simple architecture and validated only in simulation.
Data-driven approaches via Reinforcement Learning (RL) have also been applied to plasma shape control on TCV \cite{rl_ctrl3_22}, DIII-D \cite{rl_ctrl1_25}, and HL-3 \cite{rl_ctrl2_25}.
FPGA-based acceleration has also been proposed for extreme inference latencies \cite{fpga1}, but falls outside the scope of this work, which targets a standard single-core CPU.

In this paper, we present a PINO based on a modified DeepONet architecture \cite{DeepONet} as a fast surrogate for \textsc{liuqe} equilibrium reconstruction on TCV, in which a branch network conditioned on the magnetic measurements modulates the spatial basis functions learned by a trunk network.
Physics-informed regularization is enforced via automatic differentiation through the trunk path, yielding losses on $\psi$, $B_r$, $B_z$, and $j$ with LIUQE-generated targets.
The model has been compiled and deployed on the TCV PCS, where it is validated in real time against \textsc{liuqe-rt} and LIH. Closed-loop performance is assessed in simulation with the FGE solver \cite{fge1} on a variety of control strategies.

Our work makes the following distinct contributions with respect to the existing literature.
First, it demonstrates, to our knowledge, the first PINO-based equilibrium reconstruction surrogate operating at~10~kHz and deployed on the Plasma Control System (PCS) of an actual tokamak.
Second, the modular branch/trunk architecture allows the learned flux map to be evaluated at arbitrary collocation points, making the approach compatible with a variety of control strategies without retraining.

The remainder of the paper is organized as follows. Section~\ref{sec:methods} describes the model architecture and PCS deployment; Section~\ref{sec:results} presents validation and simulation results; Section~\ref{sec:conclusions} draws conclusions.

\section{Methods}
\label{sec:methods}

\subsection{Grad--Shafranov Eq. and Equilibrium Reconstruction}
\label{sec:gse}

The equilibrium of an axisymmetric magnetized plasma is governed by the GSE equation~\cite{GS1,GS2}, which expresses the balance between plasma pressure gradients and Lorentz forces in the poloidal plane $(r,z)$:
\begin{equation}
  \Delta^{*}\psi
  = -4\pi^{2}\mu_{0}\,r
  \!\left(r\,p'
  + \frac{TT'}{\mu_{0}r}\right),
  \label{eq:gse}
\end{equation}
where $\psi(r,z)$ is the poloidal magnetic flux in webers, $p'(\psi) = \mathrm{d}p/\mathrm{d}\psi$ and $TT'(\psi) = T\,\mathrm{d}T/\mathrm{d}\psi$ are the two free radial profile functions encoding the pressure gradient and the poloidal current function gradient respectively, and $\Delta^{*}$ is the Grad--Shafranov elliptic operator:
\begin{equation}
  \Delta^{*}
  = r\frac{\partial}{\partial r}
    \!\left(\frac{1}{r}\frac{\partial}{\partial r}\right)
  + \frac{\partial^{2}}{\partial z^{2}}.
  \label{eq:deltastar}
\end{equation}
The poloidal magnetic field components $B_{r}(r,z)$ and $B_{z}(r,z)$ and the toroidal plasma current density $j(r,z)$ are related to~$\psi$ through its first and second partial derivatives:
\begin{align}
  B_{r}(r,z) &= -\frac{1}{2\pi r}\frac{\partial\psi}{\partial z}, \qquad
  B_{z}(r,z)  = \phantom{-}\frac{1}{2\pi r}\frac{\partial\psi}{\partial r}, \label{eq:Bfield} \\
  j(r,z) &= -\frac{1}{2\pi\mu_0 r}
              \!\left[
              \frac{\partial^{2}\psi}{\partial r^{2}}
              +\frac{\partial^{2}\psi}{\partial z^{2}}
              \right]
              +\frac{1}{2\pi\mu_0 r^{2}}\frac{\partial\psi}{\partial r}.
  \label{eq:jtor}
\end{align}

On TCV, \textsc{liuqe} \cite{liuqe1} parametrizes the plasma current density as a linear combination of basis functions and iteratively converges to the equilibrium on the spatial grid.
\textsc{liuqe-rt} limits the number of iterations per time step and adopts simplified interpolation routines, achieving execution times of $\lesssim 0.5$~ms.
LIH implements the coarse measurement-based current-density estimate used to initialize \textsc{liuqe}'s Picard iterations: the current distribution is represented with a small set of bilinear finite elements, while the corresponding magnetic responses are obtained from precomputed geometry-dependent operators \cite{liuqe1}. It operates at~10~kHz but with significantly reduced accuracy.

\subsection{Dataset Generation}
\label{sec:dataset}

The training and validation dataset is generated by applying \textsc{liuqe} \cite{liuqe1} to all available TCV shots acquired between December~2022 and November~2025.
Shots and individual time steps are filtered using basic thresholds on plasma current and on the numerical convergence of the iterative scheme, retaining only well-converged equilibria.
One timestamp in every six is retained to reduce computational cost while preserving temporal variability; the resulting dataset covers approximately~10\,000 shots with a broad coverage of operating conditions.

The resulting dataset spans a wide variety of plasma configurations, including limiter and diverted plasmas, standard single-null configurations (ITER-Baseline-like scenarios, the most frequent), negative and positive triangularity shapes, double-null and snowflake divertors, highly elongated plasmas, and shapes associated with plasma ramp-up and ramp-down transients.
As TCV has a significantly elongated chamber, a substantial range of plasma vertical positions is also represented.
The shape distribution is not uniform: single-null scenarios predominate, reflecting the operational history of the machine, but all other configurations are sufficiently represented to support generalization.

For each retained time step, the following quantities are saved: the magnetic measurements (provided by flux loops and magnetic field probes), the full poloidal flux map~$\psi(r,z)$ on the computational grid, the corresponding magnetic field maps~$B_r(r,z)$ and~$B_z(r,z)$, and the toroidal current density map~$j(r,z)$.
The dataset is split at the shot level, assigning~80\% of shots to the training set and~20\% to the validation set.
This ensures that the validation set contains only shots not seen during training, providing an unbiased estimate of generalization performance.

\subsection{Network Architecture and Training Strategy}
\label{sec:architecture}

\begingroup


\definecolor{branchcol}{RGB}{245, 126, 20} 
\definecolor{trunkcol} {RGB}{46, 10, 90} 
\definecolor{headcol}  {RGB}{193, 58, 80}
\definecolor{outcol}   {RGB}{46, 10, 90} 
\definecolor{odotcol}  {RGB}{193, 58, 80}

\tikzset{
  netbox/.style={
    draw, rounded corners=3pt,
    minimum width=2.4cm, minimum height=0.82cm,
    align=center, font=\footnotesize\bfseries, inner sep=4pt
  },
  branchbox/.style={netbox, fill=branchcol!15, draw=branchcol!70!black,
                    text=branchcol!60!black},
  trunkbox/.style={netbox, fill=trunkcol!15,  draw=trunkcol!70!black,
                    text=trunkcol!60!black},
  headbox/.style={netbox, fill=headcol!15,   draw=headcol!70!black,
                    text=headcol!60!black},
  embbox/.style={
    draw, rounded corners=3pt,
    minimum width=1cm, minimum height=0.68cm,
    align=center, font=\footnotesize, inner sep=3pt
  },
  outbox/.style={
    draw, rounded corners=3pt,
    minimum width=1.85cm, minimum height=0.68cm,
    align=center, font=\footnotesize, inner sep=3pt,
    fill=outcol!12, draw=outcol!60!black, text=outcol!70!black
  },
  iolabel/.style={font=\footnotesize, align=center},
  odotnode/.style={
    circle, draw=odotcol, fill=white, thick,
    minimum size=0.60cm, font=\normalsize, inner sep=0pt
  },
  arr/.style    ={-{Stealth[length=5pt,width=4pt]}, thick},
  brcharr/.style={arr, color=branchcol!80!black},
  trnkarr/.style={arr, color=trunkcol!80!black},
  headarr/.style={arr, color=headcol!80!black},
  diffarr/.style={-{Stealth[length=4pt,width=3.5pt]}, thick,
                  color=outcol!80!black},
  badge/.style={rounded corners=2pt, inner sep=2.5pt, font=\tiny},
}

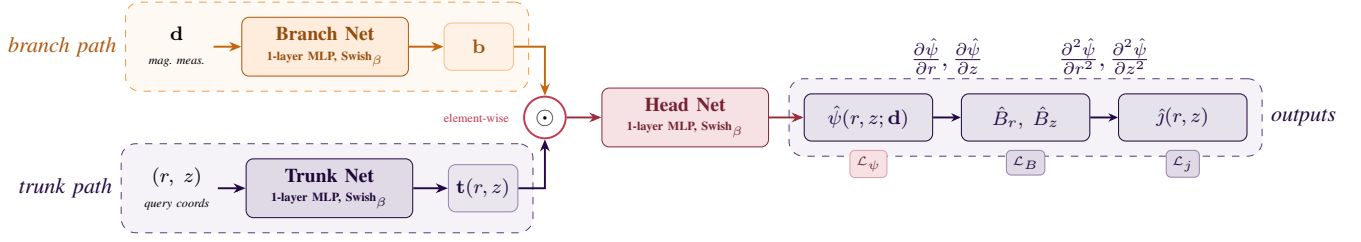
\begin{figure*}[]
\centering
\vspace{+2mm}  
\begin{adjustbox}{max width=\textwidth}

\begin{tikzpicture}[node distance=0pt]

  \node[iolabel] (d_label) {%
    $\mathbf{d}$\\[1pt]
    \textit{\tiny mag.\ meas.}};

  \node[iolabel, below=1.2cm of d_label] (rz_label) {%
    $(r,\;z)$\\[1pt]
    \textit{\tiny query coords}};

  \node[branchbox, right=.4cm of d_label] (branch) {%
    Branch Net\\[-1pt]
    \mbox{\tiny 1-layer MLP, Swish$_{\beta}$}};

  \node[trunkbox, right=.4cm of rz_label] (trunk) {%
    Trunk Net\\[-1pt]
    \mbox{\tiny 1-layer MLP, Swish$_{\beta}$}};

  \node[embbox, right=.5cm of branch,
        fill=branchcol!10, draw=branchcol!50, text=branchcol!65!black] (b_emb)
       {$\mathbf{b}$};

  \node[embbox, right=.5cm of trunk,
        fill=trunkcol!10, draw=trunkcol!50, text=trunkcol!65!black] (t_emb)
       {$\mathbf{t}(r,z)$};

  \coordinate (mid_emb) at ($(b_emb.east)!0.5!(t_emb.east)$);
  \node[odotnode, right=.1cm of mid_emb] (odot) {$\odot$};
  \node[font=\tiny, text=odotcol, left=0.08cm of odot] {element-wise};

  \node[headbox, right=.5cm of odot] (head) {%
    Head Net\\[-1pt]
    \mbox{\tiny 1-layer MLP, Swish$_{\beta}$}};

  \node[outbox, right=.5cm of head] (psi_out)
       {$\hat\psi(r,z;\mathbf{d})$};

  \node[outbox, right=0.4cm of psi_out] (Br_out)
       {$\hat{B}_r,\;\hat{B}_z$};

  \node[outbox, right=0.4cm of Br_out] (j_out)
       {$\hat{\jmath}(r,z)$};

  \coordinate (mid_psi_br) at ($(psi_out.east)!0.5!(Br_out.west)$);
  \node[font=\itshape, text=outcol!70!black,
        above=0.5cm of mid_psi_br, align=center]
       {$\tfrac{\partial\hat\psi}{\partial r},\tfrac{\partial\hat\psi}{\partial z}$};

  \coordinate (mid_br_j) at ($(Br_out.east)!0.5!(j_out.west)$);
  \node[font=\itshape, text=outcol!70!black,
        above=0.5cm of mid_br_j, align=center]
       {$\tfrac{\partial^2\hat\psi}{\partial r^2},\tfrac{\partial^2\hat\psi}{\partial z^2}$};

  \node[badge, fill=headcol!15, draw=headcol!50, text=headcol!60!black,
        below=0.12cm of psi_out] {$\mathcal{L}_\psi$};
  \node[badge, fill=outcol!15, draw=outcol!50, text=outcol!60!black,
        below=0.12cm of Br_out] {$\mathcal{L}_B$};
  \node[badge, fill=outcol!15, draw=outcol!50, text=outcol!60!black,
        below=0.12cm of j_out]  {$\mathcal{L}_j$};

  \draw[brcharr] (d_label.east)  -- (branch.west);
  \draw[trnkarr] (rz_label.east) -- (trunk.west);
  \draw[brcharr] (branch.east)   -- (b_emb.west);
  \draw[trnkarr] (trunk.east)    -- (t_emb.west);
  \draw[brcharr] (b_emb.east) -| (odot.north);
  \draw[trnkarr] (t_emb.east) -| (odot.south);
  \draw[headarr] (odot.east)  -- (head.west);
  \draw[headarr] (head.east)  -- (psi_out.west);
  \draw[diffarr] (psi_out.east) -- (Br_out.west);
  \draw[diffarr] (Br_out.east)  -- (j_out.west);

  \begin{scope}[on background layer]
    \node[rounded corners=6pt, fill=branchcol!5, draw=branchcol!80, dashed,
          fit=(d_label)(branch)(b_emb), inner sep=6pt,
          label={[font=\small\color{branchcol!70!black}]west:\textit{branch path}}] {};
    \node[rounded corners=6pt, fill=trunkcol!5, draw=trunkcol!80, dashed,
          fit=(rz_label)(trunk)(t_emb), inner sep=6pt,
          label={[font=\small\color{trunkcol!70!black}]west:\textit{trunk path}}] {};
    \node[rounded corners=6pt, fill=outcol!5, draw=outcol!80, dashed,
          fit=(psi_out)(Br_out)(j_out), inner sep=6pt,
          label={[font=\small\color{outcol!70!black}] east:\textit{outputs}}] {};
  \end{scope}

\end{tikzpicture}

\end{adjustbox}
\caption{Schematic of the proposed PINO architecture. The branch network encodes the vector of magnetic measurements~$\mathbf{d}$ into a latent embedding~$\mathbf{b}\in\mathbb{R}^{K}$. The trunk network encodes the query coordinates~$(r,z)$ into a matching embedding~$\mathbf{t}(r,z)\in\mathbb{R}^{K}$. The two embeddings are combined via element-wise multiplication and passed to a shallow regression head that produces the predicted poloidal flux~$\hat\psi(r,z;\mathbf{d})$. Magnetic field components~$\hat B_r$ and~$\hat B_z$ are obtained via automatic differentiation through the trunk path.}
\label{fig:net_arch}
\end{figure*}

\endgroup

\subsubsection{Neural Operator Foundation}

The network architecture is grounded in the DeepONet framework introduced by Lu et al. \cite{DeepONet}, which targets the learning of mappings between functional spaces, an approach referred to as operator learning.
The underlying theoretical foundation is the Universal Approximation Theorem for operators \cite{uniapprox95}, which states that any nonlinear continuous operator $G : \mathbb{R}^{m} \to \mathbb{R}^{n}$ mapping an input function~$u(\mathbf{x})$, sampled at locations~$\mathbf{x}\in\mathbb{R}^{m}$, to output values at query points~$\mathbf{y}\in\mathbb{R}^{n}$, can be approximated as
\begin{equation}
  G(u)(\mathbf{y}) \approx \sum_{k=1}^{K} b_{k}\!\left(u(\mathbf{x})\right)\,t_{k}(\mathbf{y}),
  \label{eq:deeponet}
\end{equation}
where $b(\cdot)$ and $t(\cdot)$ are the \emph{branch} and \emph{trunk} networks respectively, and $K$ is the number of latent units.
Because $\mathbf{x}$ and $\mathbf{y}$ are treated independently, the model can be trained and evaluated at arbitrary query points, improving flexibility and generalization capability.

Some prior works have applied Neural Operator or PINN-type architectures to GSE-related problems.
Refs.~\cite{gse_pinn1,bonotto24} use a DeepONet structure in which position encoding is performed on a regular grid and partial derivatives required for the physics-informed loss are estimated via finite differences.
Refs.~\cite{gse_pinn2,gse_pinn3} instead use point collocation and compute derivatives directly via automatic differentiation, but are limited either to a single fixed equilibrium or to a small set of parameter variations, and restrict collocation points to the interior of the plasma boundary.
The present approach retains the branch/trunk decomposition and automatic differentiation for derivatives, while conditioning the solution on the full vector of magnetic measurements and using LIUQE-generated targets, which allows collocation points both inside and outside the LCFS.

\subsubsection{Architecture}

The network consists of three sub-networks, as illustrated in Fig.~\ref{fig:net_arch}.

The \emph{branch network} encodes the vector of magnetic measurements~$\mathbf{d}\in\mathbb{R}^{134}$ into a latent representation:
\begin{equation}
  \mathbf{b} = \mathrm{BranchNet}(\mathbf{d}) \in \mathbb{R}^{K}.
\end{equation}
The \emph{trunk network} encodes the spatial query point~$(r,z)$:
\begin{equation}
  \mathbf{t}(r,z) = \mathrm{TrunkNet}(r,z) \in \mathbb{R}^{K}.
\end{equation}
The two embeddings are combined via element-wise multiplication, and the result is passed to a shallow \emph{head network} that produces the predicted flux: 
\begin{equation}
  \hat{\psi}(r,z;\,\mathbf{d}) = \mathrm{HeadNet}_{\psi}(\mathbf{t}(r,z)\odot\mathbf{b})
\end{equation}
This formulation modifies the original DeepONet, which uses a dot product between branch and trunk outputs, by replacing it with an element-wise product followed by a regression head.
The branch network thereby acts as a \emph{conditioner}: it encodes the current plasma scenario, captured in the magnetic measurements~$\mathbf{d}$, and modulates the spatial basis functions provided by the trunk.
All three sub-networks are implemented as single-layer MLPs with the custom activation function described below.

\subsubsection{Activation Function}

Because the physics-informed losses (introduced below) require first- and second-order partial derivatives of~$\hat\psi$ with respect to~$(r,z)$, and these derivatives must themselves be differentiated through the optimiser during training, a third-order differentiable activation function is necessary.
The Swish function \cite{swish17} satisfies this requirement and is infinitely differentiable:
\begin{align}
  \operatorname{Swish}_{\beta}(x) &= x\,\sigma(\beta x) = \frac{x}{1+e^{-\beta x}}, \\
  \frac{\mathrm{d}}{\mathrm{d}x}\operatorname{Swish}_{\beta}(x) &= \sigma(\beta x) + \beta x\,\sigma(\beta x)\bigl[1-\sigma(\beta x)\bigr]. \label{eq:der_swish}
\end{align}
where~$\boldsymbol{\beta}$ is a vector of trainable per-unit parameters that controls the sharpness of the activation.
This trainable parametrization allows a single-layer network to achieve high nonlinear expressivity without increasing depth, which is critical for inference speed.

\subsubsection{Physics-Informed Regularization}

Three loss terms are propagated jointly through the network during training.
Using Eqs.~\eqref{eq:Bfield}--\eqref{eq:jtor}, the predicted field components~$\hat B_r$,~$\hat B_z$, and~$\hat j$ are obtained by automatic differentiation of~$\hat\psi$ with respect to~$(r,z)$.
The individual loss terms are:
\begin{align}
  \mathcal{L}_{\psi} &= \frac{1}{N_{c}}\sum_{i=1}^{N_{c}}
    \bigl|\hat{\psi}(r_{i},z_{i}) - \psi_{\textsc{liuqe}}(r_{i},z_{i})\bigr|^{p} \label{eq:Lpsi}\\
  \mathcal{L}_{B} &= \frac{1}{N_{c}}\sum_{i=1}^{N_{c}}
    \bigl|\boldsymbol{\hat{B}}(r_{i},z_{i}) - \boldsymbol{B_{\textsc{liuqe}}}\bigr|^{p} \label{eq:LB}\\
  \mathcal{L}_{j} &= \frac{1}{N_{c}}\sum_{i=1}^{N_{c}}
    \bigl|\hat{j}(r_{i},z_{i}) - j_{\textsc{liuqe}}\bigr|^{p} \label{eq:Lj}
\end{align}
where $p\in\{1,2\}$ (both $\ell_1$ and $\ell_2$ norms were found to work in practice), $N_c$ is the number of collocation points per batch, and all targets are taken directly from \textsc{liuqe}.
The total training loss is the weighted sum:
\begin{equation}
  \mathcal{L}
  = \lambda_{\psi}\mathcal{L}_{\psi}
  + \lambda_{B}\mathcal{L}_{B}
  + \lambda_{j}\mathcal{L}_{j}.
  \label{eq:totalloss}
\end{equation}
The weights $\lambda_{\psi}$ and $\lambda_{B}$, with $\lambda_{\psi} = 10\,\lambda_{B}$, are kept constant throughout training, giving higher importance to direct flux accuracy.
The weight~$\lambda_{j}$ is initialized at a relatively large value and decayed exponentially during training: the current-density loss accelerates convergence in the early stages, but is tapered toward the end because $j$ is not a control target, and also to avoid overfitting to the core current-density profiles, which are ill-posed with respect to the boundary magnetic measurements and vary across scenarios.

\subsubsection{Collocation Point Sampling}

During training, query points~$(r_i,z_i)$ are drawn from a non-uniform distribution that concentrates samples near the LCFS.
This importance sampling strategy serves two purposes.
First, the LCFS and its neighbourhood are the most control-relevant region of the domain, since virtually all shape control quantities, gap distances, X-point positions, flux-surface geometry, are derived from features of~$\psi$ near the plasma boundary.
Second, because the model does not receive plasma pressure or current profiles as inputs, consistent with standard practice in real-time equilibrium reconstruction, sampling fewer points in the plasma core reduces the influence of profile-dependent reconstruction errors on the training loss.
In the absence of profile information, the network is expected to learn the most probable core profile consistent with the measurement-conditioned boundary solution.

\subsubsection{Training Procedure}

\begin{figure}[htbp]
  \centering
  \includegraphics[width=\linewidth]{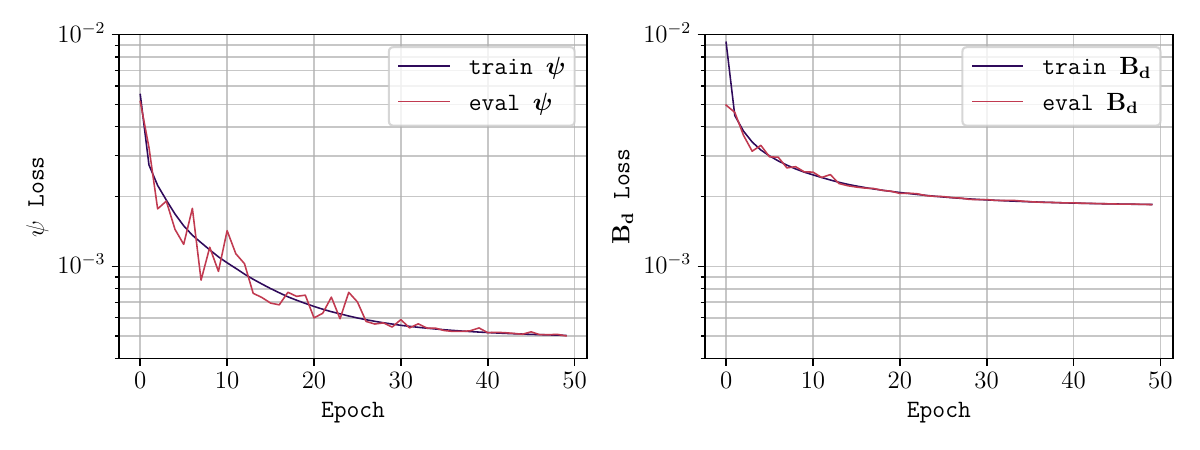}
  \caption{Training and validation loss curves for the flux loss~$\mathcal{L}_\psi$ (left) and the magnetic-field loss~$\mathcal{L}_B$ (right).}
  \label{fig:losses}
\end{figure}

The model is trained using the Adam optimiser with a batch size of~256 and a learning rate that decays logarithmically from~$5\times10^{-3}$ to~$5\times10^{-5}$ over~50 epochs.
The branch/trunk split architecture was found to converge significantly faster and to achieve the same reconstruction accuracy with fewer total trainable parameters compared to a monolithic fully-connected network with identical input/output dimensions.
Training and validation loss curves for~$\mathcal{L}_\psi$ and~$\mathcal{L}_B$ are shown in Fig.~\ref{fig:losses}.

\subsection{Implementation on the Plasma Control System}
\label{sec:implementation}

Deployment on the TCV plasma control system (PCS) \cite{tcv24ctrl} required the network to be compiled to efficient native code compatible with the real-time control infrastructure.
The model was implemented as a MATLAB function (r2019a) with code-generation compatibility, allowing the MATLAB/Simulink engine to compile it into efficient C code.
This approach guarantees modularity and interoperability with the other control system submodules.

Magnetic field components~$B_r$ and~$B_z$ are required at run time and are computed by performing a single backpropagation step through the trunk-network path, exploiting the same automatic differentiation mechanism used during training.
Profiling of the compiled code revealed that approximately~80\% of the real-time execution time was consumed by the evaluation of exponentials in the Swish activation function.
A key optimization was to cache the exponential values computed during the forward pass and reuse them during the backward pass for~$B_r$,~$B_z$ computation: because the forward and backward activations share the same sigmoid (see Eq.~\ref{eq:der_swish}), this avoids redundant exponential evaluations and reduces the total inference time by nearly a factor of two.
A further optimization, caching and reusing the trunk embedding across time steps when the set of control points does not change within a shot, was identified but deliberately not implemented, in order to preserve the modularity of the approach and allow dynamic control point selection at run time.

The module was deployed as a submodule of the full control scheme on a reserved CPU core (Intel i9) of the TCV control server, where it can run in parallel with the active control system during shots, allowing real-time speed characterization without interfering with machine operations.
Two network variants were characterized, as summarized in Table~\ref{tab:rt_perf}. Both variants operate comfortably within the~100~$\mu$s budget required for~10~kHz control.

\begin{table}
  \centering
  \caption{Deployed versions and measured reconstruction time}
  \begin{tabular}{lccccc}
    \hline\noalign{\vskip 1pt}
    Vers. & Emb.\ size & Trunk & Branch & Head & $t_{\mathrm{r}}$ ($\mu$s)\\
    \hline\noalign{\vskip 1pt}
    Small  & 32 & 32 & 48 & 48 & 78 $\pm$ 5\\
    Turbo  & 16 & 16 & 32 & 32 & 20 $\pm$ 1\\
    \hline
  \end{tabular}
  \label{tab:rt_perf}
\end{table}

\section{Results}
\label{sec:results}

\subsection{Network Reconstruction Capabilities}
\label{sec:offline_results}

\begin{figure*}[htbp]
  \centering
  \includegraphics[width=.999\textwidth]{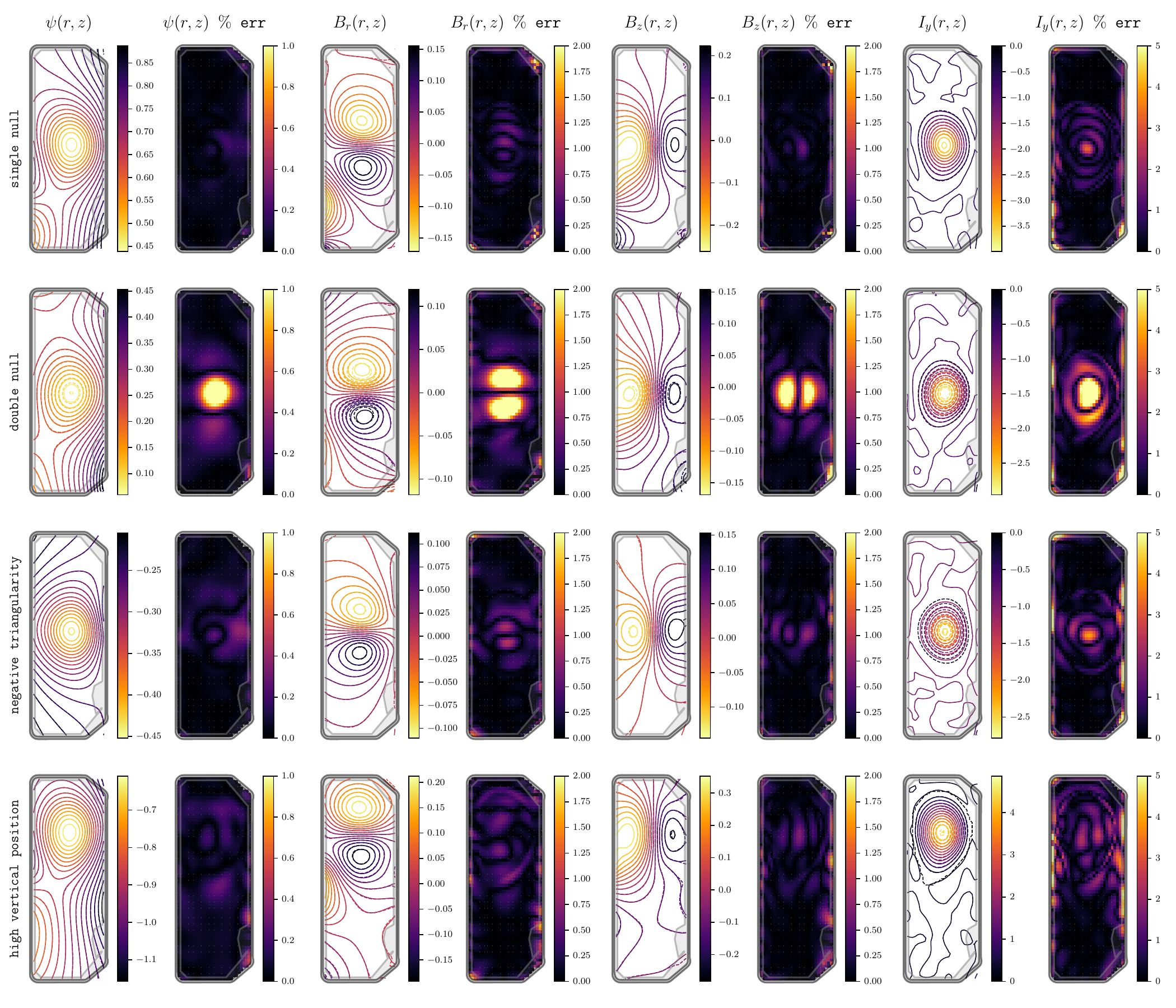}
  \caption{Reconstruction examples from the validation set for 4 distinct plasma configurations (rows). Collocation points are selected to be the entire regular grid to evaluate contours. The eight columns show, from left to right: predicted vs.\ ground-truth poloidal flux contours (PINO: solid, \textsc{liuqe}: dotted), absolute percentage error on~$\psi$ (saturated at~1\%), $B_r$ contours, percentage error on~$B_r$ (saturated at~2\%), $B_z$ contours, percentage error on~$B_z$ (saturated at~2\%), scaled toroidal current density~$\mu_0j$ contours, and percentage error on~$\mu_0j$ (saturated at~5\%). Row~1: standard single-null. Row~2: double-null. Row~3: Negative triangularity. Row~4: High vertical position.} \
  \label{fig:net_example}
\end{figure*}

Figure~\ref{fig:net_example} shows validation-set reconstructions for four plasma configurations (see caption for column layout).

The flux $\psi$ achieves the best accuracy, consistent with its dominant weight ($\lambda_\psi = 10\,\lambda_B$) in the training loss.
$B_r$ and $B_z$ are accurate across most of the domain, with somewhat higher errors near the vessel corners, regions outside the plasma vessel not covered by training collocation points.
$j$ shows the largest errors, as expected: it is not a primary control objective, its reconstruction from boundary measurements alone is ill-posed, and $\lambda_j$ is deliberately decayed to avoid overfitting core profiles.
The single-null row achieves the best accuracy across all fields, reflecting its predominance in the training set.
The double-null row shows good accuracy near the boundary but higher core errors, as the network must infer the core current-density profile from measurements rather than receiving it as input.
The remaining configurations confirm good generalization to advanced shapes and a wide range of vertical positions.

\subsection{Comparison with LIUQE-RT and LIH}
\label{sec:comparison}

\begin{figure}[htbp]
  \centering
  \includegraphics[width=\linewidth]{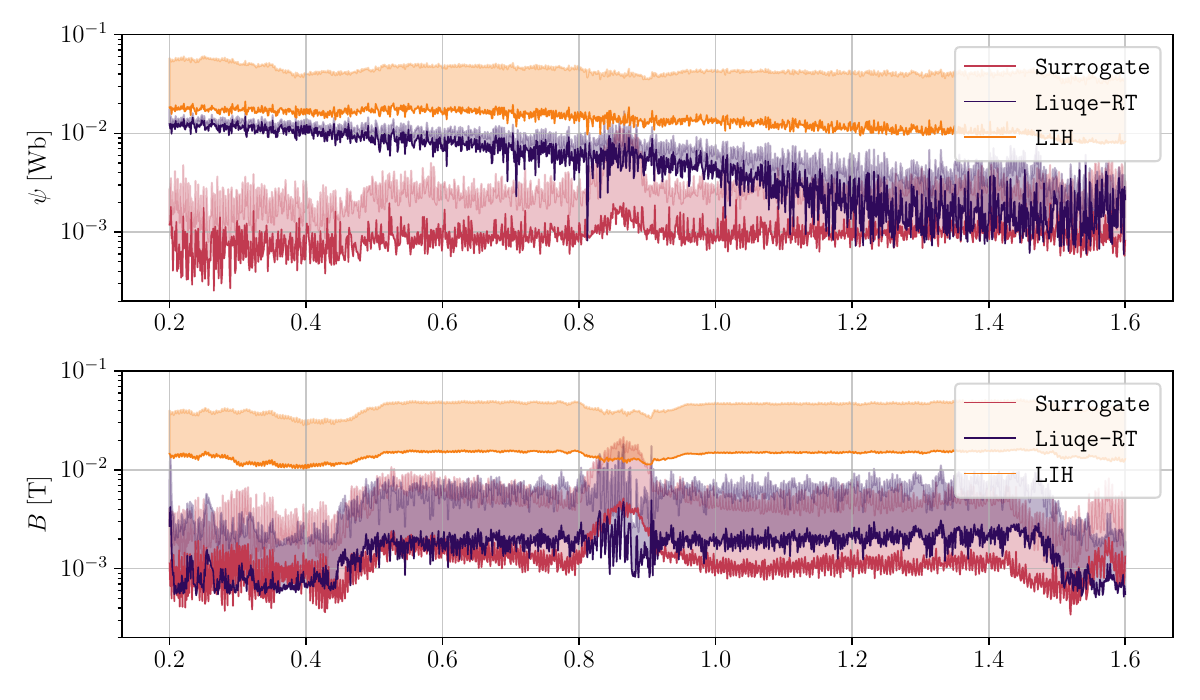}
  \caption{Reconstruction errors at 25 reference control points for a representative shot. Each observer's lower and upper traces are the mean and maximum error, respectively; the offline \textsc{liuqe} solution is the ground truth.}
  \label{fig:comparison1}
\end{figure}

The reconstruction accuracy of the PINO surrogate is compared against \textsc{liuqe-rt} and \textsc{LIH}, using the offline \textsc{liuqe} solution as the ground truth.
Figure~\ref{fig:comparison1} shows the average and maximum absolute errors across 25 reference boundary-control points from~0.2~s to~1.6~s for $\psi$ and $\sqrt{B_r^2+B_z^2}$.

Figure~\ref{fig:comparison1} compares real-time PINO with offline \textsc{liuqe-rt} and LIH results obtained for the same TCV discharge. LIH shows the largest, nearly constant error, as expected for a zeroth-iteration estimate. PINO and \textsc{liuqe-rt} have comparable accuracy, with PINO more accurate for $\psi$ during the first half of the discharge. Table~\ref{tab:observer_summary} pools the approximate flat-top intervals of 14 previously unseen shots. Both PINO variants improve the flux reconstruction over \textsc{liuqe-rt}; Small has similar field errors to \textsc{liuqe-rt}, while the faster Turbo remains substantially more accurate than LIH.

\begin{table}[t]
  \centering
  \caption{Timing and pooled flat-top reconstruction errors across 14 new shots (mean $\pm$ std over control-point/time samples).}
  \scriptsize
  \setlength{\tabcolsep}{3pt}
  \begin{tabular}{lcccc}
    \hline\noalign{\vskip 1pt}
    Observer & Timing & $\psi$ [mWb] & $B_r$ [mT] & $B_z$ [mT] \\
    \hline\noalign{\vskip 1pt}
    PINO Small & $78\pm5~\mu$s & $1.07\pm1.37$ & $1.46\pm2.36$ & $1.19\pm1.90$ \\
    PINO Turbo & $20\pm1~\mu$s & $1.95\pm2.54$ & $2.77\pm4.21$ & $2.79\pm3.65$ \\
    \textsc{liuqe-rt} & $\lesssim0.5$~ms & $6.81\pm4.25$ & $1.28\pm2.35$ & $1.04\pm1.90$ \\
    LIH & 10~kHz & $13.6\pm13.9$ & $10.6\pm14.5$ & $11.0\pm13.2$ \\
    \hline
  \end{tabular}
  \label{tab:observer_summary}
\end{table}

\subsection{Closed-Loop Control Simulations}
\label{sec:control_simulation}

The PINO surrogate is assessed in closed-loop simulations with the TCV-validated FGE free-boundary equilibrium and transport solver \cite{fge1,fge2} and three controllers of increasing complexity; solid and dashed lines denote PINO and the default observer, respectively.

\begin{figure}[htbp]
  \centering
  \includegraphics[width=\linewidth]{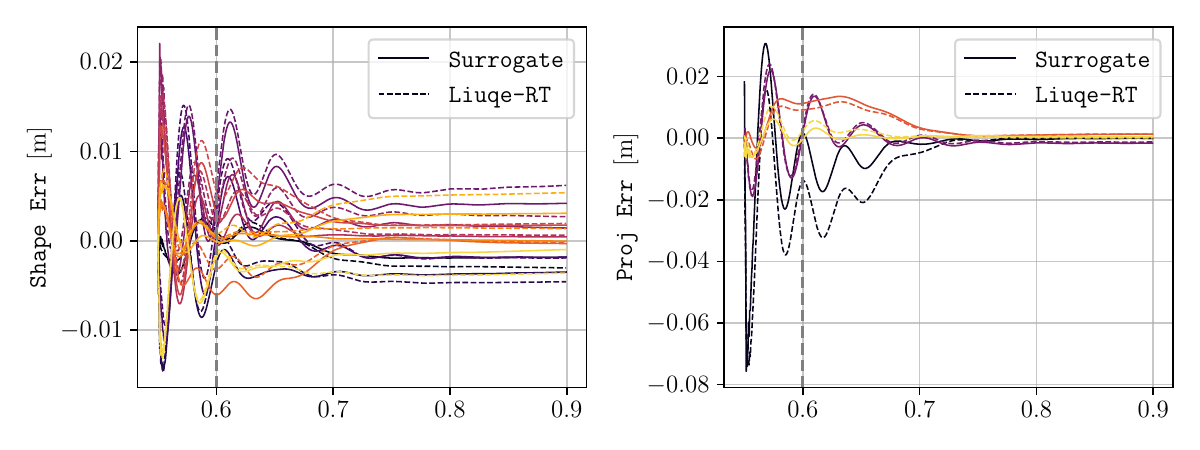}
  \caption{PID closed-loop simulation \cite{ctrl17tcv} (default: \textsc{liuqe-rt}). Left: errors at 10 boundary-control points. Right: projected errors. The dashed vertical line marks activation at~$t=0.6$~s.}
  \label{fig:pid_shape_control_simulation}
\end{figure}

\textbf{PID gap-based controller.}
The first case uses the PID-based hybrid shape controller of \cite{ctrl17tcv}, which tracks a set of gap distances on the plasma boundary and activates at~$t=0.6$~s with \textsc{liuqe-rt} as the default observer. Figure~\ref{fig:pid_shape_control_simulation} shows the errors at the 10 boundary-control points (left) and their projection in the controller basis (right). Both observers produce comparable closed-loop behaviour: all error channels converge after activation, with PINO settling marginally faster on some channels.

\begin{figure}[htbp]
  \centering
  \includegraphics[width=\linewidth]{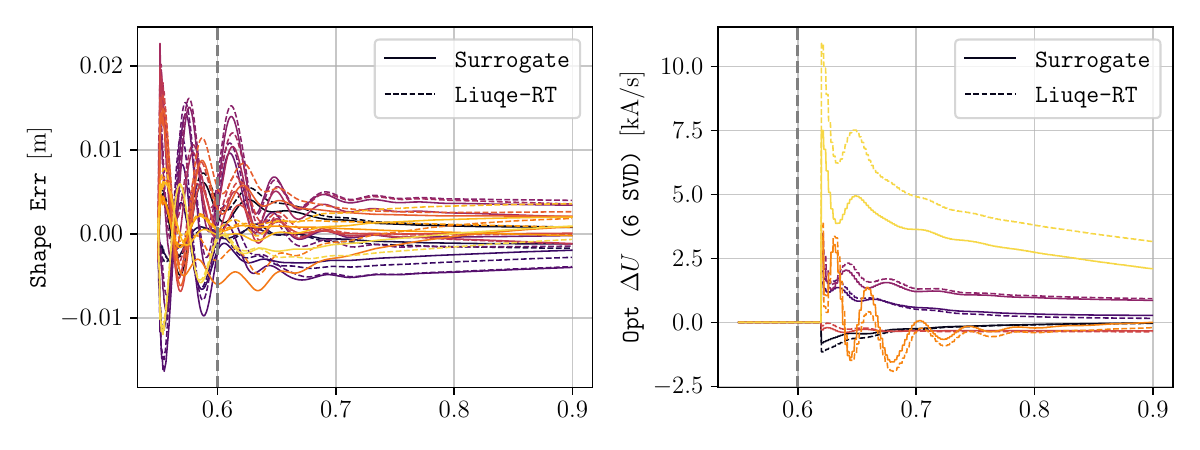}
  \caption{MPC closed-loop simulation \cite{ctrl25mpc} (default: \textsc{liuqe-rt}). Left: flux-based errors at 10 control points. Right: first six SVD components of the optimal $\Delta U$ [kA/s].}
  \label{fig:mpc_shape_control_simulation}
\end{figure}

\textbf{MPC isoflux-based controller.}
The second case uses the Model Predictive Controller of \cite{ctrl25mpc}, which tracks flux-surface targets rather than gap distances. Its QP decision variable is the optimal control increment $\Delta U$, expressed in an SVD-reduced basis of the control space. The controller activates at~$t\approx0.62$~s after Kalman-filter convergence; Fig.~\ref{fig:mpc_shape_control_simulation} shows the flux-based shape errors (left) and the first six SVD components of $\Delta U$ (right). The two observers give similar shape errors and, after the initial transient, similar optimal control increments, indicating that they drive the MPC toward the same control solution.

\begin{figure}[htbp]
  \centering
  \includegraphics[width=\linewidth]{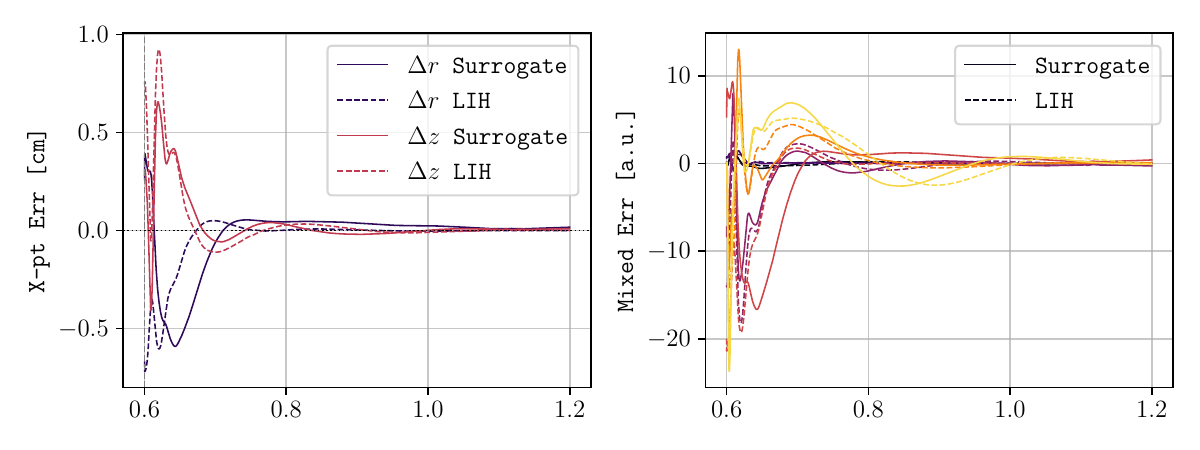}
  \caption{FMag closed-loop simulation at~10~kHz (default: LIH). Left: X-point errors $\Delta r$ and $\Delta z$ [cm]. Right: mixed control errors [a.u.].}
  \label{fig:fmag_control_simulation}
\end{figure}

\textbf{FMag mixed controller.}
The third case uses the experimental FMag controller, which simultaneously regulates plasma current, radial and vertical position, and X-point location at~10~kHz. Its default equilibrium observer is LIH rather than \textsc{liuqe-rt}; PINO replaces LIH only for X-point observation, while the position and plasma-current observers are identical in both runs. Figure~\ref{fig:fmag_control_simulation} shows the X-point position errors $\Delta r$ and $\Delta z$ (left) and the mixed control errors (right). Both configurations drive the controlled channels toward zero, with comparable X-point tracking and mixed-error evolution.

Table~\ref{tab:closed_loop_summary} reports post-activation RMS errors. For PID and MPC, the metric is pooled over the $N_c=10$ displayed control-error channels and the $N_t$ post-activation time samples: $e_{\mathrm{RMS}}=[(N_cN_t)^{-1}\sum_{i=1}^{N_c}\sum_{k=1}^{N_t}e_i(t_k)^2]^{1/2}$, where $e_i(t_k)$ is the error of channel~$i$ at time~$t_k$. For FMag, the metric is the RMS Euclidean X-point error $e_{\mathrm{RMS}}^X=[N_t^{-1}\sum_{k=1}^{N_t}(\Delta r(t_k)^2+\Delta z(t_k)^2)]^{1/2}$. PINO provides comparable control performance to the respective default observer for all three controllers.

\begin{table}[!h]
  \centering
  \caption{Closed-loop RMS error after activation.}
  \scriptsize
  \setlength{\tabcolsep}{5pt}
  \begin{tabular}{llcc}
    \hline\noalign{\vskip 1pt}
    Controller & RMS metric & PINO & Default \\
    \hline\noalign{\vskip 1pt}
    PID & Shape error [mm] & 2.75 & 3.60 (\textsc{liuqe-rt})\\
    MPC & Shape error [mm] & 2.57 & 3.03 (\textsc{liuqe-rt})\\
    FMag & X-point error [mm] & 2.06 & 2.11 (\textsc{LIH})\\
    \hline
  \end{tabular}
  \label{tab:closed_loop_summary}
\end{table}

\section{Conclusions}
\label{sec:conclusions}

We have presented a PINO surrogate for real-time plasma equilibrium reconstruction on TCV, targeting the 10~kHz bandwidth required for advanced shape control.
Based on a modified DeepONet \cite{DeepONet} with branch/trunk decomposition, it enforces physics-informed regularization via automatic differentiation using LIUQE-generated targets on $\psi$, $B_r$, $B_z$, and $j$.

Trained on around 10\,000 TCV shots and deployed on the TCV PCS, the surrogate achieves 20-80~$\mu$s inference at 25 control points, well within the 100~$\mu$s budget for 10~kHz operation.
Offline and online validation demonstrated accuracy comparable to or better than \textsc{liuqe-rt}, and closed-loop FGE simulations \cite{fge1} with three controllers of increasing complexity, a PID gap-based controller, an isoflux-based MPC, and a 10~kHz experimental mixed controller, confirmed comparable control performance in every case, establishing the viability of PINO-based equilibrium reconstruction as a practical real-time observer.

Two limitations remain.
First, the absence of plasma profile inputs leads to higher reconstruction errors in the plasma core.
Second, generalization to strongly out-of-distribution or completely new configurations is not guaranteed and should be verified before deploying the surrogate as the sole observer.

Natural extensions include \emph{pre-shot fine-tuning}, a brief fine-tuning pass before the discharge to improve accuracy for demanding target shapes; adding plasma profile inputs to reduce core reconstruction errors; and the transition to a \emph{predictive} role, forecasting plasma response to PF coil changes over a short horizon to enable model-based predictive control.

The PCS deployment reported here consists of passive parallel execution during TCV experiments, whereas all controller-driving results are obtained in software-in-the-loop FGE simulations. Real-time closed-loop testing on actual TCV shots, in which the PINO surrogate directly drives the shape controller in the PCS, is currently in preparation and will be the subject of a forthcoming contribution.

\section*{Acknowledgments}
{\footnotesize\setlength{\parskip}{0pt}\setlength{\baselineskip}{9pt}
This work has been carried out within the framework of the EUROfusion Consortium, funded by the European Union via the Euratom Research and Training Programme (Grant Agreement No 101052200, EUROfusion). Views and opinions expressed are however those of the author(s) only and do not necessarily reflect those of the European Union or the European Commission. Neither the European Union nor the European Commission can be held responsible for them. This work was also partially funded by the Italian Ministry of University and Research (MUR) under the National Recovery and Resilience Plan (PNRR), funded by the European Union -- NextGenerationEU.

}

\bibliographystyle{IEEEtran}
\bibliography{ref}

\end{document}